\documentclass[twocolumn,preprintnumbers,aps]{revtex4}%
\usepackage{graphics}
\usepackage{setspace}
\usepackage[pdftex]{graphicx}
\begin{document}
\title{Analysis of Strain Fields in Silicon Nanocrystals}
\author{D\"undar E. Y{\i}lmaz}
\email{dundar@fen.bilkent.edu.tr}
\author{Ceyhun Bulutay}
\email{bulutay@fen.bilkent.edu.tr}
\affiliation{Department of Physics, Bilkent University, Ankara, 06800, Turkey}
\author{Tahir \c{C}a\u{g}{\i}n}
\email{cagin@chemail.tamu.edu}
\affiliation{Texas A$\&$M University, Artie McFerrin Department of Chemical Engineering,\\
Jack E. Brown Engineering Building, 3122 TAMU, College Station, TX 77843-3122, USA}
\begin{abstract}
Strain has a crucial effect on the optical and electronic properties of nanostructures.
We calculate the atomistic strain distribution in silicon nanocrystals up to a diameter of 3.2~nm embedded in an amorphous silicon dioxide matrix. 
A seemingly conflicting picture arises when the strain field is expressed in terms of bond lengths versus volumetric strain. 
The strain profile in either case shows uniform behavior in the core, however it becomes nonuniform within 2-3~\AA~distance
to the nanocrystal surface: tensile for bond lengths whereas compressive for volumetric strain. We reconsile their coexistence
by an atomistic strain analysis.
\end{abstract}
\maketitle 
The low dimensional forms of silicon embedded in silica have strong potential as an optical material.\cite{nanosi}
Such heterogeneous structures inherently introduce the strain as a degree of freedom for optimizing their opto-electronic properties.
It was realized earlier that strain can be utilized to improve the carrier mobility in bulk silicon 
based structures.\cite{manasevit} This trend has been rapidly transcribed to lower dimesional structures, starting with 
two-dimensional silicon structures.\cite{people} Recently for silicon nanowires, there have been a 
number of attempts to tailor their optical properties through manipulating strain.\cite{lyons,hong}
Futhermore, recent studies have revealed that the strain 
can become the major factor restricting the crystallization of the nanolayers.\cite{zachariasPRB,hernandez}
It depends on several factors, most important of which are the lattice mismatch between 
the constituents, size of the NCs, and the growth conditions, such as the details of the growth 
procedure.\cite{zach99}
In summary, for improving the optical  and electronic properties of nanocrystals (NCs), the strain engineering  
has become an effective tool to be exploited.\cite{thean,peng,smith}  
A critical challenge in this regard is to analyze the strain state of the Si NCs embedded in silica.

The close relations between strain and optical or electronic properties in Si NCs have very recently
become the center of attention.\cite{peng,hadji07,ossicini} There still remains much to be done in order to understand strain in 
nanostructures at the atomistic level. As pioneered by Tsu \textit{et al.} Raman spectroscopy can be an effective experimental tool for determining the strain state of the Si NCs.\cite{tsu} Specifically, recent Raman studies
reported that the Si NCs may be under a thermal residual strain and this can be reduced by overall annealing at high 
temperatures\cite{zach99} or by local laser annealing.\cite{arguirov}
Due to small density difference between Si NC and the surrounding a-SiO$_2$, a limited information can 
be gathered about its structure using transmission electron microscopy (TEM) or 
even high resolution TEM techniques.\cite{coffin} Especially, molecular dynamics
simulations with realistic interaction potentials present an opportunity, by providing more detailed critical information then the best imaging techniques currently available and clarify the analysis of experimental results.  
Along this direction, previously \cite{dundar} we focused 
on Si-Si bond length distribution and reported that Si-Si bond lengths are stretched upto 
3$\%$ just below the surface of Si NCs embedded in amorphous SiO$_2$ which has also been very recently 
confirmed.\cite{ippolito} 

%\section{Theory and Computation}
In this Letter, we analyze the volumetric and bond length strain distributions in Si NCs, in particular demonstrate 
that both compressive volumetric strain and tensile bond length strain coexist within the same Si NC. 
We accomplish this by performing trajectory analysis on model samples (with ca. 5000 atoms) simulated via molecular 
dynamics using a reliable and accurate as well as reactive force field.\cite{adri}  The simulation details are 
similar to our previous work,\cite{dundar} except the way we construct the Si NC in glass matrix. 
Instead of deleting all glass atoms within a predetermined radius, we remove the glass atoms after rigorously defining the 
surface of the nanocrystal through the Delaunay triangulation method.\cite{geompack} In this way, 
we have constructed NCs embedded in glass matrix with diameters ranging from 2.2~nm to 3.2~nm without introducing 
built-in strain to the system. In this diameter range we observe similar trends in strain, volumetric strain, and  
bond length distribution etc., therefore, we present only the figures of the system for a typical NC of 
radius 2.6~nm. 

%\section{Results}
In the language of geometry, strain is defined through an affine transformation that maps the undeformed state to deformed state, which is called deformation gradient. Several methods to derive discrete form of deformation gradient from atomic positions are reported.\cite{gullett,pryor,curtin} In the method proposed by Pryor \textit{et al.}, the atomistic strain tensor is derived from local transformation matrix that transforms nearest neighbors of a certain atom from its undeformed state to the deformed one. From the MD simulations, using positions of NC atoms, we first extract each atom’s displacement vector from its undeformed site which is determined by positioning an ideal tetrahedron to the local environment. Using these displacement vectors, we construct  deformation matrix and derive the atomistic strain tensor from this local deformation tensor.\cite{pryor}  The first invariant of strain tensor corresponds to the hydrostatic strain.\cite{elasticity} As an alternative measure to hydrostatic strain, we calculate volumetric strain by considering volume change of a tetrahedron from its undeformed counterpart. A third measure as we have used in our previous report,\cite{dundar} is the bond length strains.

\begin{figure}
\includegraphics{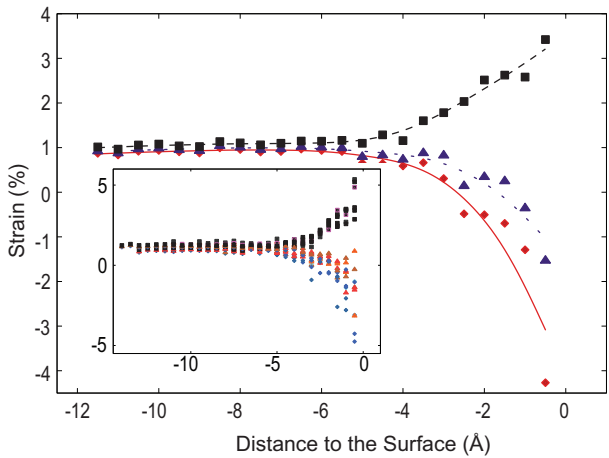}
\caption{(Color online) 
Variation of, Si-Si bond lengths (squares), hydrostatic strain  (diamonds), 
and the volumetric strain (triangles)  as a function of distance to nanocrystal surface (see text). Dashed, dotted and solid lines are guides to the eye for the respective data set. 
The data for 2.6~nm diameter NC is used. Inset: Other NC diameters ranging from 
2.2~nm to 3.2~nm are also shown.}
\label{fig:strainfig}
\end{figure}

\begin{figure}
\includegraphics{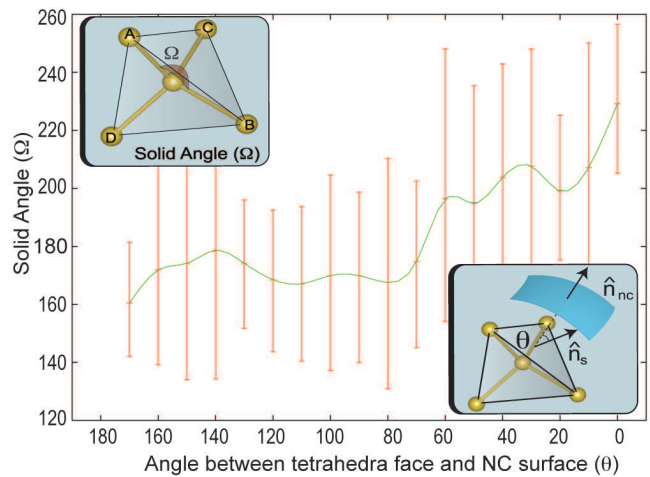}
\caption{(Color online) 
Dependence of solid angle subtended by tetrahedron face to the angle between tetrahedron face and nanocrystal surface. 
Illustration of solid angle subtended by tetrahedron face (top left inset) and the angle between tetrahedron 
face and NC surface (bottom right inset).}
\label{fig:solidangle}
\end{figure}

To verify our results we have calculated strain distribution in NC region for all mentioned measures. We have plotted all three of them in Fig.~\ref{fig:strainfig}.
The results of volumetric strain are very close to hydrostatic strain which is the trace of strain tensor calculated with aferomentioned technique.\cite{pryor} 
In these results, we observe a net compressive behavior of strain just under the surface and 
a uniform tensile strain of about 1$\%$ at the core of NC. Si-Si bonds are stretched by about 1$\%$  in the core region in agreement with the hydrostatic and volumetric strains, however, just under the surface, Si-Si bonds are stretched up to 3$\%$  where hydrostatic and volumetric strain 
results indicate compressive strain state. The bond-stretch in Si-Si bonds due to oxidation has been shown 
earlier by us using molecular dynamics simulations\cite{dundar} which was also confirmed by 
other approaches.\cite{ippolito} 
%---------------------------figure-1-------------------------------------------------------------
%----------------------------------------------
%---------------------------------------------------------------------------------------------
Occurrence of compressive volumetric strain and stretched bond lengths in the same outer region may initially seem 
contradicting. However, stretching of bonds does not imply that the system is under tensile 
hydrostatic strain as well. Consider a tetrahedron formed by a Si atom and its four Si neighbors (A, B, C, D) as
shown in upper inset of Fig.~\ref{fig:solidangle}. In the ideal case, the solid angle 
($\Omega$) subtended by each triangular face of this tetrahedron should be equal to $180^\circ$. 
Under a uniform deformation, bond lengths will also be stretched, while the solid angles remain 
unchanged. However, under a nonuniform deformation, the change in three solid angles causes a decrease 
in the volume of the tetrahedron while increasing or preserving the bond lengths. 
Hence, a combination of stretched bond lengths with deformed solid angles may end up with 
an overall reduction of the volume of the tetrahedron.
This explains the coexistance of compressive volumetric strain and stretched bond lengths at 
the region just below the surface of NCs. 

To better visualize the nature of the deformation of the Si NCs, we consider the orientational 
variation of the solid angles of the tetrahedral planes. As illustrated in the lower inset of 
Fig.~\ref{fig:solidangle}, the two important directions are the unit normal 
($\hat{n}_{\mbox{\begin{scriptsize}S\end{scriptsize}}}$) 
of the tetrahedron face subtending the solid angle under consideration, and the local 
outward surface normal ($\hat{n}_{\mbox{\begin{scriptsize}NC\end{scriptsize}}}$) 
of the NC. It is clearly seen from Fig.~\ref{fig:solidangle} that solid angles subtended by 
tetrahedra faces oriented outward to the NC surface are increased up to $220^\circ$, whereas those 
facing inward to the NC core are decreased down to $160^\circ$. This dependence is a 
clear evidence of how oxidation affects strain distribution close to the interface.
%--------------------------------figure-2------------------------------------------------------------------
%----------------------------------------------------------------------------------------------------------

\begin{figure}
\includegraphics{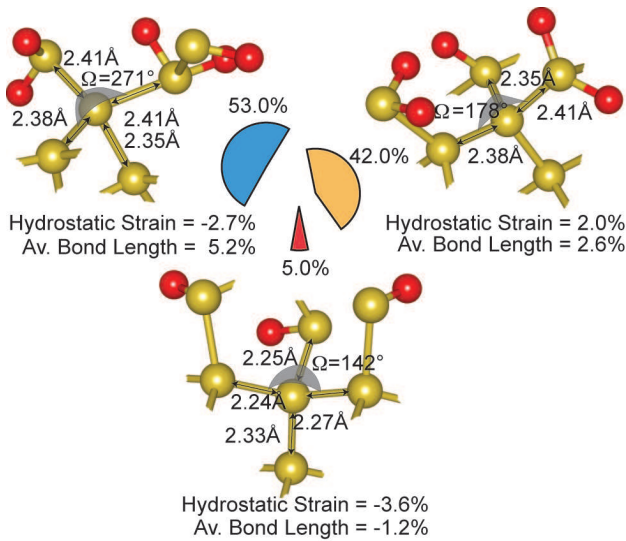}
\caption{(Color online) 
Illustrations of oxidation effects on strain in three categories with their percentage of occurrences: 
Si-Si bonds are stretched and system is under compressive strain (upper left). Si-Si bonds are stretched 
and system is under tensile strain (upper right). Si-Si bonds are shortened and the Si atom at the center 
is under compressive strain (bottom). Large spheres (gold) and small spheres (red) represents 
Si and O atoms, respectively.}
\label{fig:categories}
\end{figure}

To further quantify the atomistic strain in the highly critical region within 3~\AA~distance 
to the interface, we classify the average bond length and hydrostatic strain behaviors into three 
categories. Figure~\ref{fig:categories} displays the percentage as well as the 
bonding details of each category.
%----------------------figure-3 ----------------
%--------------------------------------------------------------------------------
In top-left, we illustrate most common type with a share of  53.0$\%$ which is responsible for the 
opposite behavior in Fig.~\ref{fig:strainfig} where average bond lengths of center Si atoms to 
its four nearest neighbors are stretched but net atomistic strain at this atom is compressive. 
In this case solid angles facing toward the oxide region is increased to $270^\circ$ due to 
oxygen bonds of Si neighbors. Although these oxygen bonds stretched Si-Si bonds to $2.41$~\AA, 
net strain on center Si atom is -2.7$\%$. In the top-right part of the Fig.~\ref{fig:categories} 
we illustrate second most often case with a percentage of 42.0$\%$, where average bond 
lengths and atomistic strain show similar 
behavior; bond lengths are stretched and net hydrostatic strain is tensile. In this case oxidation 
somewhat uniformly deforms the bonds so that solid angles are still around $180^\circ$ which is the value for the unstrained case. 
Finally, as shown at bottom of Fig.~\ref{fig:categories}, a very small percentage of atoms (5.0$\%$) 
in the region beneath the surface have shortened bond lengths and compressive atomistic strain. 

%\section{Conclusion}
In summary, we study the strain state of Si NCs in silica matrix with diameters in 2 to 3.2~nm. 
The structure is assumed to be free from any thermal built-in strains. The core region 
of the NC is observed to be under a uniform 1\% tensile strain, where both bond length and 
volumetric strain measures are in agreement.
However, towards the NC interface, while the Si-Si bonds become more stretched, the 
hydrostatic strain changes in the compressive direction. In the interpretation of the indirect 
strain measurements eg. from spectroscopy, this dual character needs to be taken into consideration.
We explain these two behaviors using the solid angle deformation of the tetrahedral-bonded 
Si atoms, and demonstrate that it is ultimately caused by the oxygen atoms at the interface. 
An equally important finding is that the overall strain profile within the Si NCs is quite 
nonuniform. As very recently emphasized, within the context of centrosymmetric 
materials, like silicon, such strain gradients locally break the inversion symmetry and 
may lead to profound physical consequences.\cite{majdoub}

%\section{Acknowledgements}
This work has been supported by the Turkish Scientific and Technical Council T\"UB\.ITAK with the
Project No. 106T048 and by the European FP6 Project SEMINANO with the Contract No. 
NMP4 CT2004 505285. The visit of Tahir
\c{C}a\u{g}{\i}n to Bilkent University was facilitated by the T\"UB\.ITAK
B\.IDEB-2221 program. TC also acknowledges the support of NSF-IGERT.
%========================================================================

\newpage
%---------------------figures---------------------------------------------------

\end{document}